\documentclass[prl,twocolumn,showpacs]{revtex4}

\usepackage{amsmath,amsfonts,amssymb,bm}
\usepackage{dcolumn}
\usepackage[final]{graphicx}
\usepackage{comment}
\usepackage{bm}

\includecomment{pdffigure}

\begin{document}

\bibliographystyle{apsrev}	

\title{Tomography of particle plasmon fields from electron energy loss spectroscopy}

\author{Anton H\"orl}
\author{Andreas Tr\"ugler}
\author{Ulrich Hohenester}%
  \email{ulrich.hohenester@uni-graz.at}%

\affiliation{Institut f\"ur Physik,
  Karl--Franzens--Universit\"at Graz, Universit\"atsplatz 5,
  8010 Graz, Austria}

\date{August 2, 2013}

\begin{abstract}

We theoretically investigate electron energy loss spectroscopy (EELS) of metallic nanoparticles in the optical frequency domain.  Using a quasistatic approximation scheme together with a plasmon eigenmode expansion, we show that EELS can be rephrased in terms of a tomography problem.  For selected single and coupled nanoparticles we extract the three-dimensional plasmon fields from a collection of rotated EELS maps.  Our results pave the way for a fully three-dimensional plasmon-field tomography and establish EELS as a quantitative measurement device for plasmonics.

\end{abstract}



\maketitle


Electron energy loss spectroscopy (EELS) has emerged as an ideal tool for the study of surface plasmon polaritons (SPPs) and particle plasmons~\cite{garcia:10}.  For SPPs, electrons with kinetic energies of a few to hundreds of keV penetrate through a metal film and excite surface and bulk plasmons, whose resonance frequencies can be directly extracted from the energy loss spectra~\cite{ritchie:57,powell:59}.  By raster scanning the electron beam over a plasmonic nanoparticle, one can extract both the resonances and field maps of the particle plasmons~\cite{bosman:07,nelayah:07}.  This technique has been extensively used in recent years to map out the plasmon modes of nanotriangles~\cite{nelayah:07,nelayah:09,schaffer:10}, nanorods~\cite{bosman:07,schaffer.prb:09,nicoletti:11,rossouw:13}, nanodisks~\cite{schmidt:12}, nanocubes~\cite{mazzucco:12}, nanoholes~\cite{sigle:09}, and coupled nanoparticles~\cite{chu:09,ngom:09,koh:09,koh:11}.

Despite its success and widespread application, the interpretation of plasmonic EELS data remains unclear.  In \cite{garcia:08} the authors speculated that EELS renders the photonic local density of states (LDOS), a quantity of immense importance in nano optics~\cite{novotny:06}, but the interpretation was put in question in \cite{hohenester.prl:09}.  A detailed comparison between LDOS and EELS was recently given in \cite{boudarham:12}, where the authors provided an intuitive interpretation of different measurement schemes in terms of an eigenmode expansion.  It should be noted that the controversy only concerns the interpretation, whereas the theoretical description of EELS maps is well established~\cite{garcia:10} and very good agreement between experiment and simulation has been achieved~\cite{nelayah:07,schaffer.prb:09,schmidt:12,mazzucco:12}.

In this paper we challenge the interpretation of EELS maps of plasmonic nanoparticles, and rephrase the problem in terms of a tomography scheme.  For sufficiently small nanoparticles, where the quasistatic approximation can be employed, we expand the particle fields in terms of \textit{plasmonic eigenmodes}\/~\cite{ouyang:89,mayergoyz:05,boudarham:12} and the EELS signal becomes a simple spatial average along the electron propagation direction.  We show at the example of single and coupled nanorods that the extraction of plasmon fields from EELS data can be reduced to an inverse Radon transformation, which is at the heart of most modern computer tomography algorithms~\cite{herman:80}.  Otherwise the field extraction can be formulated in terms of an inverse problem which can be solved by optimization techniques.  


\begin{figure}[b]
\centerline{\includegraphics[width=0.85\columnwidth]{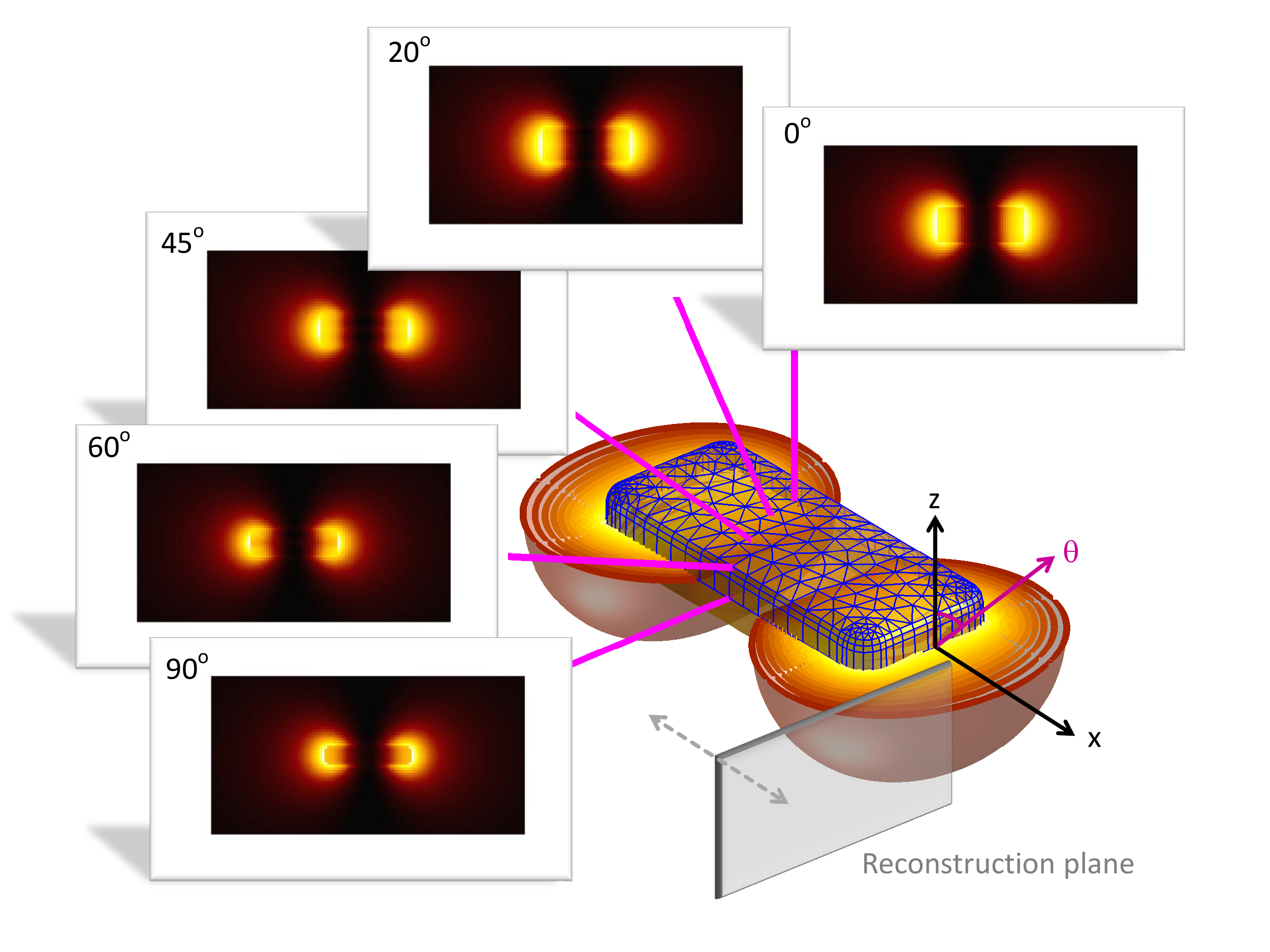}}
\caption{(Color online) Schematics of EELS tomography.  An electron beam is raster-scanned over a metallic nanoparticle, and EELS maps are recorded for different rotation angles $\theta$.  The main panel shows the isosurface and contour lines for the modulus of the dipolar surface plasmon potential, and the insets report the different EELS maps.  From the complete collection of maps one can reconstruct the plasmon fields, as described in text (positions of reconstruction planes used in Figs.~2 and 3 are indicated in main panel).}
\end{figure}

\begin{figure}
\centerline{\includegraphics[width=0.9\columnwidth]{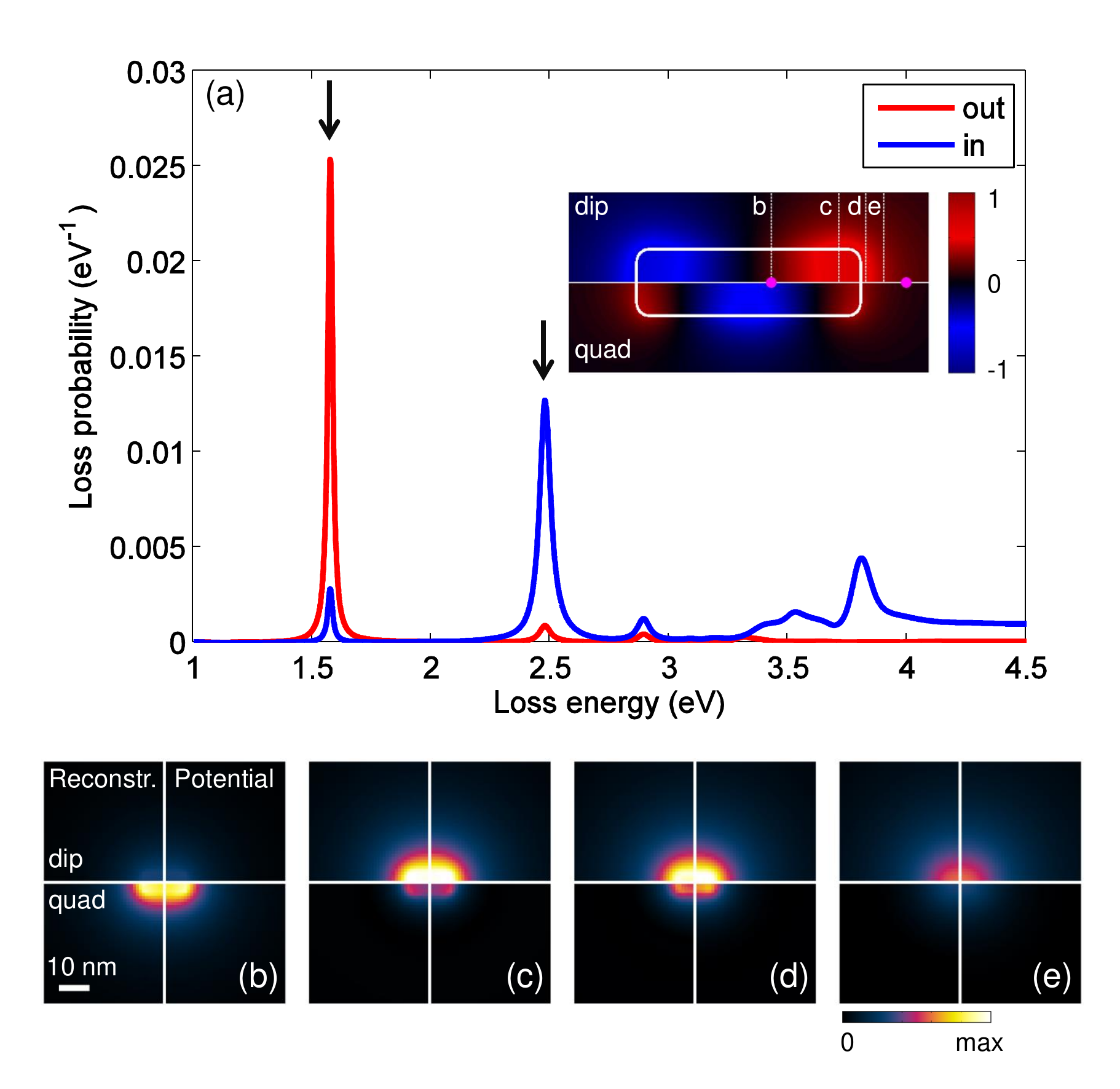}}
\caption{(Color online).  (a) EELS spectrum for silver nanorod with dimensions of $50\times 15\times 7$ nm$^3$ and for the two beam positions indicated with circles in the inset.  The inset also reports the potential maps for the dipole and quadrupole mode at $z=0$.  The dashed lines indicate the positions of the planes where the potentials are reconstructed from the collection of EELS maps.  (b--e) Potential maps reconstructed from EELS maps (left panels) and potential maps (right) for dipole mode (upper panels) and quadrupole mode (lower panel).  In the simulations we assume a kinetic electron energy of 200 keV and use a dielectric constant of $1.6$ for the embedding medium.
}
\end{figure}

\begin{figure}
\centerline{\includegraphics[width=0.9\columnwidth]{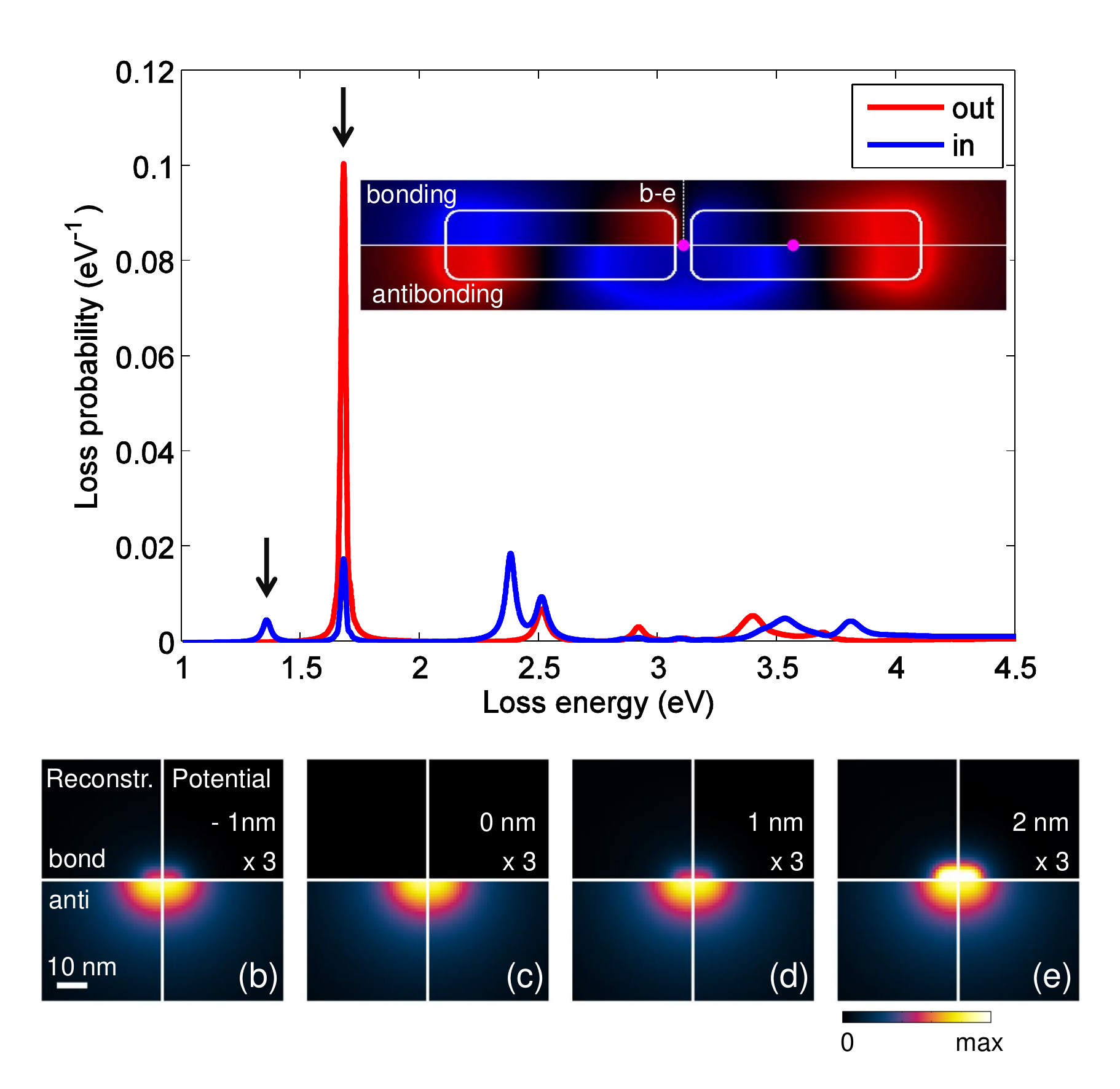}}
\caption{(Color online).  Same as Fig.~2 but for coupled nanorods.  The particle and simulation parameters are the same as those given in the caption of Fig.~2, the gap distance between the nanoparticles is 5 nm.  In the inset we report the potentials for the bonding and antibonding modes.  (b-e) Reconstructed and true potential maps at different $x$-positions, reported in the panels, as measured with respect to the gap center.  For clarity, the potentials for the bonding mode are multiplied by a factor of three.
}
\end{figure}

\textit{EELS simulation}.---Electron energy loss is a two-step process, where the electron first excites a surface plasmon and, in turn, the electron has to perform work against the induced surface plasmon field.  The energy loss becomes \cite{garcia:10,ritchie:57}
\begin{equation}\label{eq:loss}
  \Delta E=e\int \bm v\cdot\bm E_{\rm ind}[\bm r_e(t),t]\,dt=\int_0^\infty \hbar\omega
  \Gamma_{\rm EELS}(\omega)\,d\omega\,,
\end{equation}
where $-e$ and $\bm v$ are the charge and velocity of the electron, respectively, and $\bm E_{\rm ind}$ is the electric field of the surface plasmon evaluated at the electron positions.  In the second expression of Eq.~\eqref{eq:loss} we have spectrally decomposed the different loss contributions and have introduced the loss probability $\Gamma_{\rm EELS}$.  A similar expression can also be obtained from a fully quantum-mechanical description scheme \cite{garcia:10}.
For nanoparticles much smaller than the wavelength of light one can employ the quasistatic limit by keeping only the scalar potential and performing the static limit for the Green functions, while retaining the full frequency dependence for the material permittivities \cite{garcia:10}.  We are then led to \cite{garcia:97,garcia:10}
\begin{eqnarray}\label{eq:gammaeels}
  &&\Gamma_{\rm EELS}(\bm R_0,\omega)=-\frac{e^2}{\pi\hbar v^2}
  \int_{-\infty}^\infty dz\,dz'\,\\
  &&\qquad\times\,\Im m\left[
  e^{-i\omega z/v}G_{\rm ind}(\bm r_e,\bm r_e',\omega)
  e^{i\omega z'/v}\right]\,dzdz'\nonumber
\end{eqnarray}
for the loss probability.  Here $G_{\rm ind}$ is the Green function in the quasistatic limit that describes the response of the metallic nanoparticle \cite{garcia:10,hohenester.prl:09}.  We next introduce \textit{plasmonic eigenmodes}\/~\cite{ouyang:89,mayergoyz:05,boudarham:12} defined through
\begin{equation}\label{eq:eigen}
  \int_{\partial\Omega}\frac{\partial G(\bm s,\bm s')}{\partial n}\,\sigma_k(\bm s')\,da'=
  \lambda_k\sigma_k(\bm s)\,,
\end{equation}
where $\lambda_k$ and $\sigma_k(\bm s)$ denote the plasmonic eigenvalues and eigenmodes, respectively, and $\partial G/\partial n$ is the derivative of the Green function of an unbounded medium with respect to the outer surface normal.  The eigenmodes are orthogonal in the sense $\int \sigma_k(\bm s)G(\bm s,\bm s')\sigma_{k'}(\bm s)\,dada'=\delta_{kk'}$ and can be chosen real \cite{ouyang:89,mayergoyz:05}.  Let $\phi_k(\bm r)=\int_{\partial\Omega}G(\bm r,\bm s)\sigma_k(\bm s)\,da$ denote the potential of the $k$'th eigenmode.  The induced Green function can then be decomposed into these eigenmodes according to \cite{boudarham:12}
\begin{equation}\label{eq:gind}
  G_{\rm ind}(\bm r,\bm r')=-\sum_k\frac{\lambda_k\pm 2\pi}{\Lambda+\lambda_k}\,
  \phi_k(\bm r)\phi_k(\bm r')\,\frac 1{\varepsilon(\bm r')}
\end{equation}
with $\Lambda=2\pi(\varepsilon_1-\varepsilon_2)/(\varepsilon_1+\varepsilon_2)$ and $\varepsilon_1$ and $\varepsilon_2$ being the dielectric functions inside and outside the particle, respectively.  The plus and minus sign correspond to the situations where $\bm r'$ lies outside or inside the particle.  
Inserting Eq.~\eqref{eq:gind} into the loss probability of Eq.~\eqref{eq:gammaeels}, we obtain for an electron trajectory that does not penetrate the particle the final result
\begin{eqnarray}\label{eq:modeloss}
  &&\Gamma_{\rm EELS}^{\rm out}(\bm R_0,\omega)=-\frac{e^2}{\pi\hbar v^2\varepsilon_2}\\
  &&\qquad\times\sum_k 
  \Im m\left(\frac{\lambda_k+2\pi}{\Lambda+\lambda_k}\right)
  \left|\int_{-\infty}^\infty e^{i\omega z/v}\phi_k(\bm r)\,dz\right|^2\,.\nonumber
\end{eqnarray}

This expression, which has been previously derived in \cite{boudarham:12}, forms the starting point for our following analysis.  At a plasmon resonance, defined through $\Re e[\Lambda(\omega)+\lambda_k]=0$, the resonance term in Eq.~\eqref{eq:modeloss} becomes large and its contribution can dominate the total loss probability.  Let us assume for the moment that $\omega z/v\ll 1$, such that the EELS probability for the single, dominant mode reduces to
\begin{equation}\label{eq:sinogram}
  \Gamma_{{\rm EELS},\theta}^{\rm out}(\bm R_0,\omega)\sim
  \Bigl|\mathcal{R}_\theta[\phi_k(\bm r)]\Bigr|^2\,.
\end{equation}
Here $\mathcal{R}_\theta$ is the \textit{Radon transformation}\/ \cite{herman:80,midgley:03} that performs a line integration of $\phi_k(\bm r)$ along the $z$-direction.  We have included in Eq.~\eqref{eq:sinogram} an angle $\theta$ that accounts for a possible rotation of the integration axis, as schematically depicted in Fig.~1.  A collection of Radon transformations for a complete set of rotation angles is conveniently called a \textit{sinogram} \cite{midgley:03}.  The projection-slice theorem then states that one can uniquely reconstruct the original function from the sinogram.  Eq.~\eqref{eq:sinogram} differs from a normal sinogram in that $\Gamma_{\rm EELS}$ depends on the square of the Radon transforms, which leads to a sign ambiguity in the sinogram.  In the following we first analyze a situation where the sign ambiguity can be ignored, and we will discuss the more general situation further below.

\textit{Results}.---We first consider the setup depicted in Fig.~1, where an electron beam is raster scanned over a single nanorod and the EELS maps are recorded for different loss energies $\hbar\omega$ and rotation angles $\theta$.  In panel (a) of Fig.~2 we show the simulated EELS spectrum for the electron beam positions shown in the inset.  We use a dielectric function for silver~\cite{johnson:72} and employ the MNPBEM toolbox~\cite{hohenester.cpc:12} for the solution of the full Maxwell equations (without the quasistatic approximation).  At low loss energies one observes two peaks which can be attributed to the dipolar and quadrupolar plasmon modes.  Owing to the symmetry of the modes, an electron propagating along $z$ always passes through regions where $\phi_k(\bm r)$ is either solely positive or negative, which allows us to perform the inverse Radon transformation in Eq.~\eqref{eq:sinogram}.  Results are reported in panels (d,e), showing almost perfect agreement between the reconstructed potentials and $\phi_k(\bm r)$, apart from the potential sign that cannot be reconstructed from the EELS data.  This is an encouraging finding, considering that our EELS maps are obtained from the solutions of the full Maxwell equations.

In Fig.~3 we show EELS maps for coupled nanoparticles, which have received considerable interest in recent years \cite{hohenester.prl:09,chu:09,ngom:09,koh:09,koh:11}, partially due to their importance for surface enhanced Raman scattering (SERS) \cite{khan:06,mirsaleh:12}.  Inside the gap region the EELS signal becomes zero for the bonding mode and maximal for the antibonding mode, as discussed in detail in Ref.~\cite{hohenester.prl:09}.  However, from the reconstructed potential maps one observes a significant variation of the bonding potential along $x$, indicating a strong electric field in the gap region, contrary to the antibonding mode which has an only weak dependence along $x$.  Thus, although ``being blind to hot spots'' \cite{zabala:97,hohenester.prl:09} EELS tomography even allows to reconstruct the complete field distribution inside the gap region.

The situation becomes more complicated when the electron passes through the metallic nanoparticle, and the induced Green function in Eq.~\eqref{eq:gind} has to be separated into contributions where the electron is either inside or outside the metallic particle.  Inside the metal the electron becomes efficiently screened by free electrons through the $\varepsilon^{-1}$ term.  To a good approximation, we can ignore this contribution and approximate the EELS probability by
\begin{equation}\label{eq:sinogramin}
  \Gamma_{{\rm EELS},\theta}(\bm R_0,\omega)\sim
  \bigl(\mathcal{R}_\theta[\phi_k(\bm r)]\bigr)\bigl(\mathcal{R}_\theta[\phi_k^{\rm out}(\bm r)]\bigr)\,,
\end{equation}
where $\phi_k^{\rm out}(\bm r)$ is the potential that is artificially set to zero inside the particle.  In Eq.~\eqref{eq:sinogramin} it is no longer possible to perform an inverse Radon transformation to reconstruct the plasmon potential, and we have to proceed in a different manner.   First, we introduce a cost function that measures the distance between the computed EELS probabilities and those computed from Eq.~\eqref{eq:sinogramin}.  Let $f_0$ denote the EELS probabilities for all impact parameters and rotation angles, and $f[\phi_k(\bm r)]$ the corresponding probabilities computed from Eq.~\eqref{eq:sinogramin}.  In a second step we then determine, starting from some reasonable initial guess, those potentials that minimize the cost function $J=\frac 12\left|f_0-f[\phi_k(\bm r)]\right|^2$ using a nonlinear conjugate gradient method~\cite{roy:01}.  In most cases the initial guess for the potentials was not overly critical and the minimization algorithm converged after a few iterations.  Panels (b,c) in Fig.~2 report the reconstructed potentials and $\phi_k(\bm r)$ for electrons penetrating through the metallic nanoparticle, and we observe again very good agreement. 

Having established a numerical optimization scheme for the potential through minimization of the cost function, we can also rephrase the EELS tomography problem of Eqs.~(\ref{eq:modeloss},\ref{eq:sinogram}) in a way that appears better suited for experimental implementation and that can be also employed for more complicated structures.  To this end, we first note that the source for the potential $\phi_k(\bm r)$ is the charge distribution $\sigma_k(\bm s)$ of the eigenmodes, and one can reconstruct equally well the surface charge distribution or the potential.  We next rewrite Eq.~\eqref{eq:modeloss} in the form
\begin{equation}\label{eq:sinogramsig}
  \Gamma_{{\rm EELS},\theta}^{\rm out}(\bm R_0,\omega)=
  \sum_k C_k(\omega)\left|\int\phi_{\bm R_0,\theta}^*(\bm s)\sigma_k(\bm s)\,da\right|^2\,,
\end{equation}
where $\phi_{\bm R_0,\theta}(\bm s)=-(e/v)\int_{-\infty}^\infty G(\bm s,\bm r_e)e^{i\omega z_e/v}\,dz_e$ is the potential of the electron propagating along $\bm r_e$, with direction $\theta$ and impact parameter $\bm R_0$, and the form of $C_k(\omega)$ follows directly from the comparison with Eq.~\eqref{eq:modeloss}.  Equation~\eqref{eq:sinogramsig} allows for the reconstruction of $\sigma_k(\bm s)$, which can be approximated by boundary elements (as used in our simulation approach \cite{hohenester.cpc:12}) or some freeform surface functions such as non-uniform splines, provided that the nanoparticle surface is known \cite{midgley:03}.  In what follows, we again set $\omega/v\approx 0$.

\begin{figure}[t]
\centerline{\includegraphics[width=\columnwidth]{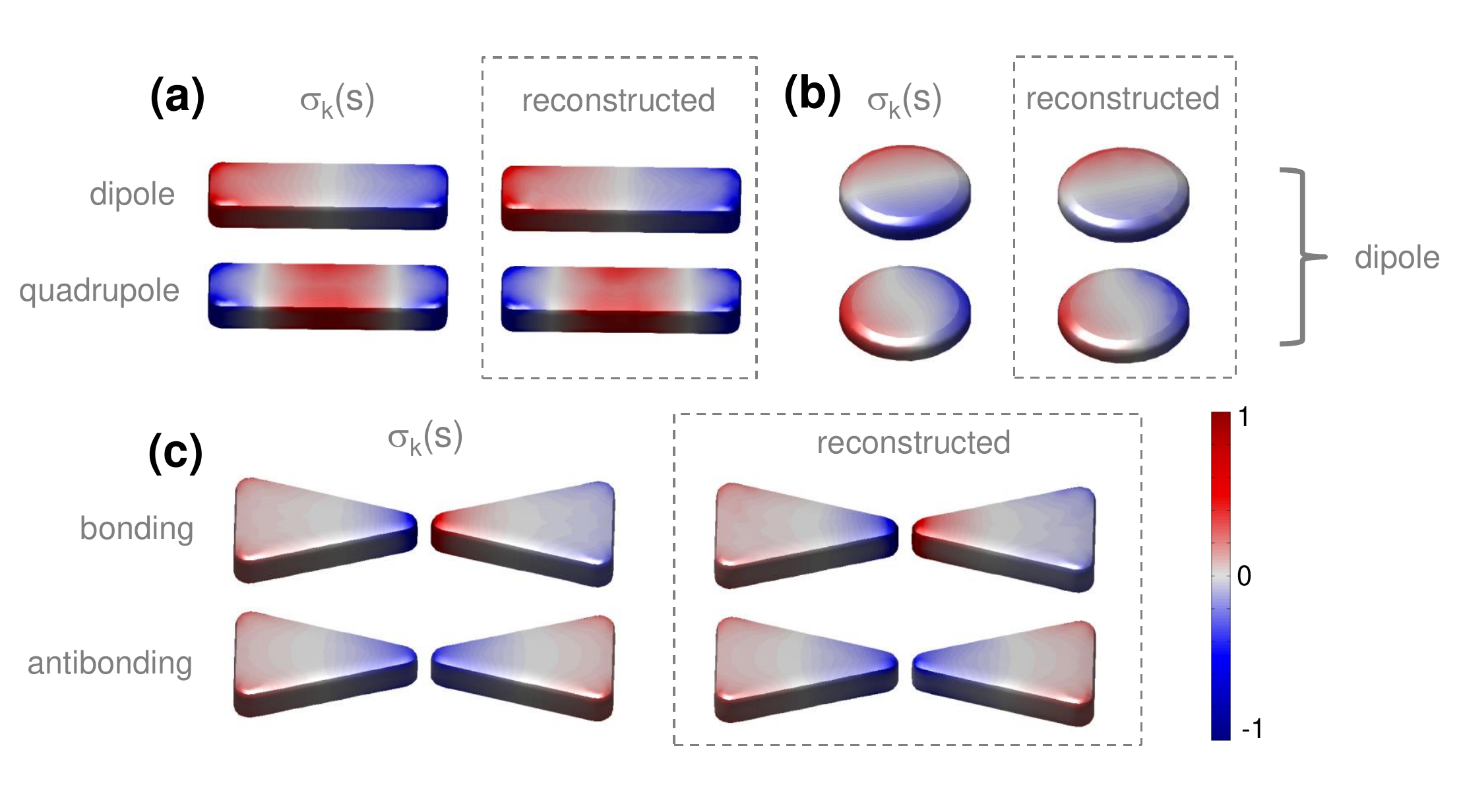}}
\caption{(Color online) Surface charge distribution $\sigma_k(\bm s)$ of eigenmodes (left) and surface charge distribution reconstructed from the EELS maps (right), using Eq.~\eqref{eq:sinogramsig}, for (a) nanorod, (b) nanodisk, and (c) bowtie geometry.  Surface charge distribution is given in arbitrary units.  For the reconstruction of $\sigma_k$ we consider EELS maps with a resolution of $40\times 40$ pixels, and use 45 rotation angles within $\theta\in[0,90^\circ]$ (similar results were obtained with only 10 rotation angles).
}
\end{figure}

Figure 4 shows for a number of particle shapes the reconstruction based on Eq.~\eqref{eq:sinogramsig}.  In all cases we used for the initial guess a mode profile with proper symmetry, whereas other details turned out to be unimportant.  Panel (a) reports $\sigma_k(\bm s)$ (left) and the reconstructed surface charge distributions (right) for the dipolar and quadrupolar nanorod modes, which are in very good agreement.  In panel (b) we show results for a disk-shaped particles with two degenerate eigenmodes.  For the reconstruction, we keep in Eq.~\eqref{eq:sinogramsig} two modes with identical coefficients $C_k$, and ensure that, because of symmetry, the charge distributions of these modes are identical but rotated by 90 degrees with respect to each other.  Again the optimization procedure comes up with the correct modes.  We emphasize that a similar approach could be used for modes that are energetically close to each other, although in this case the coefficients $C_k$ are different and the optimization should include EELS maps for different loss energies.  Finally, Fig.~4(c) shows the bonding and antibonding mode distributions for a bowtie geometry, demonstrating that our approach can be also applied to more complicated structures.

In Fig.~5 we compare for the nanorod the true and reconstructed potentials along the line (e) shown in the inset of Fig.~2(a) [$z=0$].  We observe that the quasistatic potential and the potentials reconstructed from the EELS maps, through either the Radon transformation [Eq.~\eqref{eq:sinogram}] or the surface charges of Eq.~\eqref{eq:sinogramsig}, are in good agreement, demonstrating the quantitative measurement capability of our approach.  The comparison with the retarded potentials is complicated by the fact that there exists no clear eigenmode concept for the full Maxwell equations, and we thus have to proceed in a different manner.  In the figure we show the modulus of the induced potentials for a plane-wave excitation (we use an incidence angle of 45 degrees where both dipolar and quadrupolar modes can be excited).  Good agreement between the solutions of the quasistatic and full Maxwell equations is found, with only small deviations at larger positions, attributed to the different excitation conditions and/or retardation effects not included in the quasistatic solutions.  

\begin{figure}
\centerline{\includegraphics[width=0.9\columnwidth]{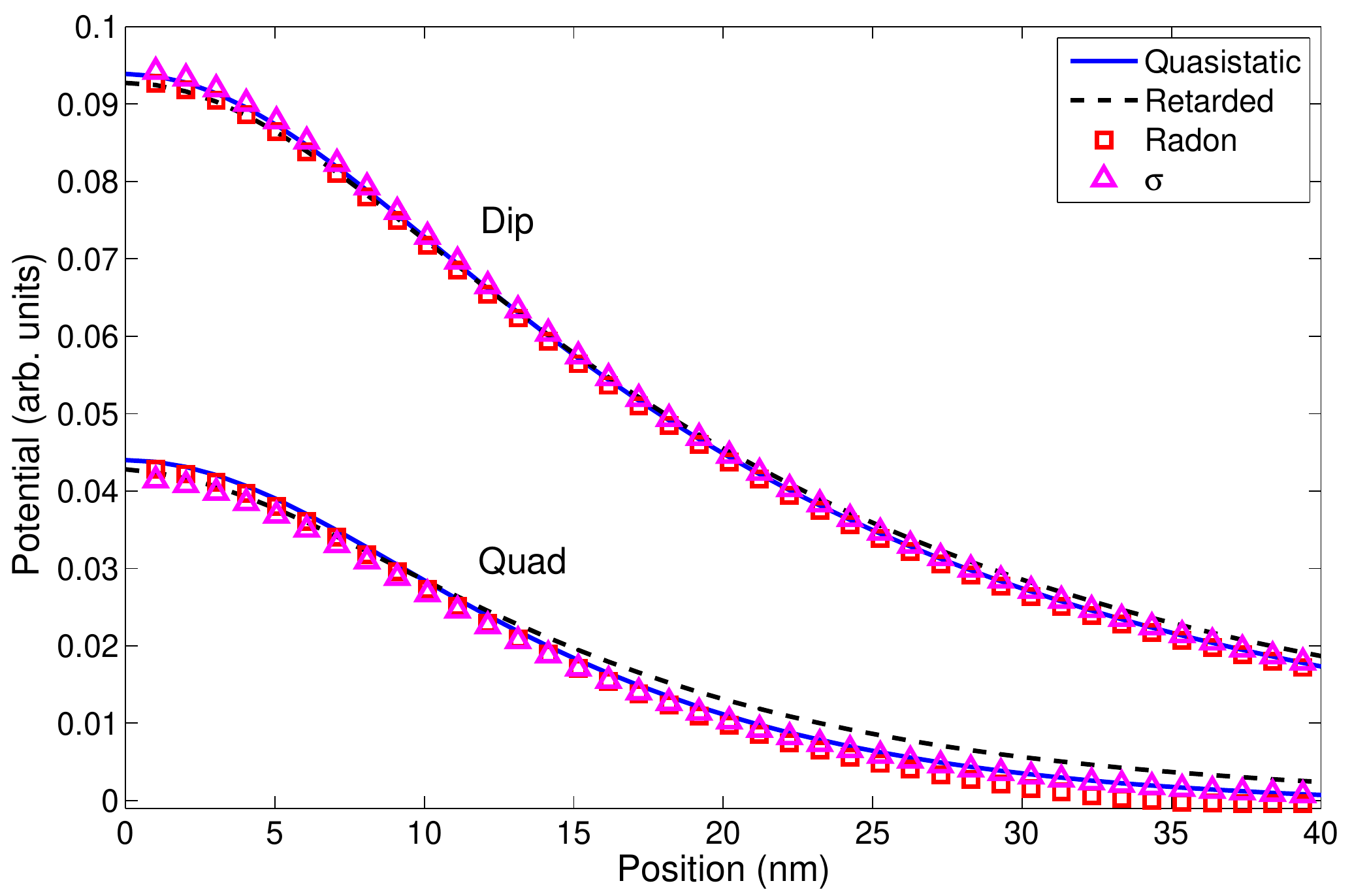}}
\caption{(Color online) Eigenmode potential $\phi_k(\bm r)$ (solid line) and reconstructed potentials along the line (e) shown in the inset of Fig.~2(a) [$z=0$].  For the reconstruction we either use the inverse Radon transformation of Eq.~\eqref{eq:sinogram} [square symbols] or the surface charge decomposition of Eq.~\eqref{eq:sinogramsig} [triangles].  The square modulus of the retarded scalar potential (scaled by an arbitrary factor) is computed for a planewave excitation, as described in text.}
\end{figure}

There are several reasons why Eq.~\eqref{eq:sinogramsig} is advantageous in comparison to Eq.~\eqref{eq:sinogram}.  First, while $\sigma_k(\bm s)$ can typically be represented by a few tens to hundreds of boundary elements or parameters, the EELS maps for different rotation angles provide a much larger data set, thus making the optimization procedure for the reconstruction a highly overdetermined problem.  The reason for this overdetermination is the two-dimensional nature of the surface charge distribution, whereas the potential, which is uniquely determined by $\sigma_k(\bm s)$, can be measured in the entire three-dimensional space.  For the reconstruction of $\sigma_k(\bm s)$ one can thus even discard trajectories where the electrons pass through the nanoparticle, which are anyhow problematic in experiment because of the electron attenuation within the metal.  The inverse Radon transformation additionally requires a large field of view, to properly include the far-reaching components of the dipolar or multipolar surface plasmon fields, in contrast to Eq.~\eqref{eq:sinogramsig} that can be restricted to significantly smaller regions.  Consideration of finite wavenumbers $\omega/v$ naturally enters the framework of Eq.~\eqref{eq:sinogramsig}, in the spirit of diffraction tomography~\cite{bronstein:02}, although in this work we have neglected for simplicity such wavenumber effects.  Finally, effects of substrates or layers supporting the nanoparticles can be included in our approach by replacing in Eq.~\eqref{eq:eigen} and in the definition of $\phi_{\bm R_0,\theta}(\bm s)$ the Green function of an unbounded medium by that including substrate or layer effects.  The main limitations of our tomography scheme are probably the quasistatic approximation, which restricts the scheme to sufficiently small particles, and the high degree of pre-knowledge needed for the surface charge reconstruction (homogeneous dielectric function of particle, surface charge distributions as only source for plasmonic fields).

\textit{Acknowledgment}.---We are grateful to Gerald Kothleitner, Toni Uusim\"aki, Franz Schmidt, Harald Ditlbacher, and Joachim Krenn for most helpful discussions.  This work has been supported by the Austrian Science Fund FWF under project P24511--N26 and the SFB NextLite.

\end{document}